\documentstyle[12pt]{article}
\begin{document}
\begin{flushright}
{\normalsize
December 1996}
\end{flushright}
\vskip 0.2in
\begin{center}
{\Large {Thermal Masses and Equilibrium Rates in the Quark Gluon Phase}}
\vskip 0.2in
{\normalsize Jan-e Alam$^a$, Pradip Roy$^a$, Sourav Sarkar$^a$,
Sibaji Raha$^b$ and Bikash Sinha$^{a,c}$}
\end{center}
\vskip 0.2in
\noindent
{\small {\it a) Variable Energy Cyclotron Centre,
     1/AF Bidhan Nagar, Calcutta 700 064, India}}
\vskip 0.15in
\noindent
{\small {\it b) Bose Institute,
           93/1, A. P. C. Road, Calcutta 700 009,
           India}}
\vskip 0.15in
\noindent
{\small {\it c) Saha Institute of Nuclear Physics,
           1/AF Bidhan Nagar, Calcutta 700 064,
           India}}

\addtolength{\baselineskip}{0.5\baselineskip} %wide line spacing
\parindent=20pt
\vskip 0.4in
\noindent
\centerline{\bf {\normalsize Abstract}}
\vskip 0.15in

We apply the momentum integrated Boltzmann transport equation 
to study 
 the time evolution of various quark flavours in the central region
 of ultra-relativistic heavy ion collisions. 
The effects of thermal masses
for quarks and gluons are incorporated to take into account 
the in-medium properties of
these ingredients of the putative quark gluon plasma. 
We find
that even under very optimistic conditions,
complete chemical equilibration in the quark gluon plasma
appears unlikely.

\vskip 0.4in
\noindent
% PACS number(s) : 25.75.+r; 12.38.Mh; 24.60.Ky; 24.85.+p
\newpage
Quantum  Chromodynamics  (QCD), the theory of
strong interactions, predicts that at very high temperature
and/or density, the colourless hadronic matter dissolves into
its coloured constituents, the quarks and gluons, such that
the bulk properties of strongly interacting matter are
governed by these degrees  of  freedom.  Such a locally
colour-deconfined phase  of  matter  is  known  as  Quark Gluon
Plasma (QGP). It is expected that the temperature and density
achievable in ultrarelativistic collisions of heavy ions is
favourable for QGP formation, although transiently.

Many aspects of this transition,{\it e.g.}, the order of the
phase transition, the value of the critical temperature,
signals of the transition, thermodynamic equilibration
, are still uncertain. So far,
most of the works assume thermodynamic equilibrium throughout the
evolution history, after an initial proper time  $\tau_i$  ($\sim
1$ fm/c). Thus the thermodynamic quantities like pressure, entropy
and temperature have direct meaning with respect to an equilibrated
system. It has however been recently shown that the approach to
equilibrium in ultrarelativistic heavy ion collisions proceeds
through a succession of many time scales$^{1,2}$. In
particular, the dominance of the $gg$ cross section over the $qg$
or $qq$ cross sections$^{3}$ was argued to imply that the
gluons equilibrate among themsleves appreciably earlier than the
whole system of quarks and gluons. In a recent work$^1$ ,
we showed that the time scale of {\it
kinetic} equilibration for the light quarks ($u,d$) is $\sim 1$
fm/c; for massive quarks, it increases with the mass to rather
large values. The significance of these considerations for QGP
diagnostics has been discussed in references 4 and 5; the emission
of particles from the pre-equilibrium era ($\tau_g \le \tau  \le
\tau_{th}$), $\tau_{th}$ denoting the time for full thermodynamic
equilibrium, may indeed populate the invariant mass or $p_T$ windows
thought to be relevant for signals from the thermalised QGP.

In this letter, we study the {\it chemical} evolution of the quarks
in the system from the epoch $\tau_g$ ( the proper time when gluons
thermalise) onward. Chemical equilibration has very important
implications for the signal of QGP formation$^{4,5,6}$ . It
would be ideal to study this problem in the QCD based kinetic theory and
attempts along these lines have been made recently$^7$ . Such
calculations, appealing as they are, rely heavily on rather involved
numerical simulations and also suffer from dependence on parameters
needed to mimic the non-perturbative effects. We propose, as in ref. 1
, a scenario which retains the essence of the
kinetic theory to the following extent : the gluons thermalise {\it
completely} at a proper time $\tau_g$ earlier than the quarks. The
gluons carry about half of the total momentum and the quarks carry a
tiny fraction. We restrict ourselves to the  central  rapidity  region
where the number of valence quarks is assumed to be negligibly small.
We shall, however, relax this condition when we look at the situation
at SPS energies.

In  order  to  treat  the  equilibration  of the species one must
follow the microscopic evolution of the phase space  distribution
function  $f(x^\mu,p^\mu)$, governed  by the Boltzmann
transport equation :
\begin{equation}
\hat  L\lbrace
f_q\rbrace  =  \hat C \lbrace f_q\rbrace
\end{equation}
where $\hat L$
is the Liouville operator and $\hat C$ the collision operator,
the subscript $q$ denoting a quark flavour.
The Boltzmann   equation   is   a  non-linear,
integro-differential equation for the  phase  space  distribution
function  of  the  particles. For
our present purpose, however, the problem can be addressed in a
rather simple manner without losing much information on the time
scales for chemical equilibrium.

The  decoupling of the relic particles in
the early universe was studied$^{8,9}$ by integrating the Boltzmann
equation over the momentum of the particle to obtain an equation
for the evolution of number density. We use a similar approach to
look at the number density evolution of the partons in
ultrarelativistic collisions of heavy ions; the major difference
between the two cases is that the expansion dynamics in the early
universe ({\it big bang\/}) is governed by the Hubble expansion$^9$
 whereas in the case of heavy ion collision ({\it mini
bang\/}), the expansion dynamics is assumed to be governed by the
boost invariant (Bjorken) solution of relativistic hydrodynamics$^{10}$
. The equation for the number density evolution then reads
\begin{equation}
\frac{dn_q}{d\tau} + \frac{n_q}{\tau}=
\frac{g_q}{(2\pi)^3}\,\int\lbrace\hat C\rbrace
\frac{d^3p}{E}
\end{equation}
where  $n_q$  is  non-equilibrium quark
density, $\tau$ the proper time and $g_q$ the statistical
degeneracy of the quarks of flavour $q$.

Clearly, the time evolution of $n_q$ would be largely governed by
the various processes determining the collision term in eq. (2). The most
important contribution to this is from gluon fusion from
($gg\rightarrow q\bar q$).

Since we have considered complete equilibration for gluons, 
reactions like $gg \leftrightarrow ggg$ need
not be included$^1$
; the gluon number density at each instant
is uniquely determined by the temperature of the system. However
the gluons in a finite temperature system acquire a sizable thermal mass
may qualify as real excitations$^{11,12}$ . Thus the gluon decay
process $(g \rightarrow q\bar q)$ and its inverse is expected to play a
major role; in particular it has been argued in the literature$^{11}$
that for small quark masses $(m_q^{eff}/T<<1)$, the gluon decay dominates
over the fusion process. We must therefore take the gluon decay into account
while evaluating the collision term. We ignore the quantum effects as the
temperature is expected to be very high but include relativistic effects
for the same reason. Thus gluon are described by the relativistic
Maxwell-Boltzman statisics, $f_g(E)=g_g/(2\pi)^3\,e^{-E/T}$.
Then eq.(2) can be written as:
\begin{equation}
\frac{dn_q}{d\tau}
=-\frac{n_q}{\tau}-\left(R_{gg\leftrightarrow q\bar q}(T)
+R_{g\leftrightarrow q\bar q}(T)\right)
\frac{(n_q^2-n_{eq}^2)}{n_{eq}^2}
\end{equation}
where
$n_{eq}(T)=g_q/(2\pi)^3\int{d^3p/(e^{E/T}+1)}$, is
the equilibrium density. Eq.(3) is a particular form of the Riccati
equation; there is no general closed form solution of this equation.

In deriving eq.(3) we have assumed T (or CP) invariance, {\it i.e.},
$|M|_{gg\rightarrow q\bar q}=|M|_{q\bar q\rightarrow gg}$
and $|M|_{g\rightarrow q\bar q}=|M|_{q\bar q\rightarrow g}$.
$R_{gg\rightarrow q\bar q}(T)$ is the quark  production  rate  per
unit four volume by the reaction $gg\rightarrow  q\bar q$ and
$R_{g\rightarrow q\bar q}$ gives the same for the gluon decay$^{11,12}$,
$g\rightarrow q\bar q$ . The thermal quark masses
have been taken into account using $m_q^{eff}(T)=
\sqrt{m_{current}^2+m_{th}^2}$
in all the rates. In the present context in a chemically non-equilibrated 
scenario $m_{th}$ is given by$^{13}$
\begin{equation}
m_{th}^2=(1+\frac{r_q}{2})\frac{g^2T^2}{9}
\end{equation}
where $g$ is the colour charge, $r_q\equiv n_q/n_{eq}$.
The full derivation of eq.(3) has been omitted
here for the sake of brevity; it will be reported elsewhere$^{14}$ .
Eq.(3)  is a very convenient form of the Boltzmann transport
equation with all the relevant features for our
present purpose. The first term on the right hand side of the equation
represents the ``dilution'' of the density due to one dimensional
expansion, the  destruction of the quarks is proportional to the
annihilation rates of $q\bar q$ and destruction is balanced by creation
through the inverse reaction when $n_q=n_{eq}$. The last term
represents the change in quark density due to gluon decay and its
inverse process, arising from the thermal gluon mass. Note that there is
no overcounting as the phase space integrations involved in the gluon
fusion and the gluon decay channels are over different domains. For a
detailed discussion see ref. 11.

To solve eq.(3) for $n_q(\tau)$ we need to know $T(\tau)$, which
is obtained by solving the hydrodynamic equation
\begin{equation}
\frac{d\epsilon}{d\tau}+\frac{\epsilon+P}{\tau}=0
\end{equation}
where $P$ is the pressure and $\epsilon$ the energy density. An
equation of state relating $P$ and $\epsilon$ can be parametrized in
the form
\begin{equation}
P=c_{s}^{2}\epsilon
\end{equation}
Then, for the non-equilibrium configuration, we can write
\begin{equation}
\epsilon=\left[ar_g+br_q(\tau)\right]T^4
\end{equation}
where $a=(8\pi^2)/45\,c_{s}^{-2}$ and
$b=(7\pi^2)/60\,N_f\,c_{s}^{-2}$.
The medium effects enter
eq. (7) through the thermal mass of the
quarks; the higher the temperature, the higher the thermal mass. Therefore
the equation of state may be parametrized as $P=c_{s}^{2}(T)\epsilon$, where
$c_{s}^{-2}=\left[\frac{T}{g_{eff}(T)}\,\frac{dg_{eff}}{dT} + 3\right]$
with $g_{eff}(T) = (\epsilon+P)/(2\pi^2 T^4)/45$. This would amount to
making $a$ and $b$ in eq. (7) functions of $T$. For the cases considered
here, such effects are negligibly small.

In chemical equilibrium,
$r_q$ for all species should be unity
and in correspondence with our premise, we take $r_g$ =1
all through. ${N_f}$ is the {\it effective} number of
{\it massless} flavours; appearance of quark masses  amounts to
reducing the actual number of flavours$^{15}$ . For the
present, we confine our attention only to the three lightest flavours,
$u,d$ and $s$. The difference between the $u(d)$ quarks and $s$ quarks arises
only in terms of their current mass. We have verified that because of the high
initial temperature and the corresponding large thermal masses, this amounts
to a very small change in the effective $N_f$.

Equations (5) and (7) together yield,
\begin{equation}
T^3\tau=\left[\frac{a+br_q(\tau_g)}{a+br_q(\tau)}
\right]^{3/4} T_g^3\tau_g
\end{equation}
which is the cooling law in the present non-equilibrium scenario.

To  solve  eq.(3), we need to specify the initial conditions.
There is no unique prescription for evaluating them at the present state
of the art. Bir\'o {\it et al.} estimated$^{16}$ the initial
parton density from the
HIJING algorithm$^{17}$ . On the other hand, Shuryak and Xiong$^{5}$,
advocated, as in ref. 1, the use of "perturbative" QCD(pQCD),
{\it i.e.}, the
deep inelastic structure functions for the estimation of the initial quark
density. Given the importance of the issue, we have, in the present work,
used both these estimates.

In the pQCD approach, the
initial  values  of  $n_q(\tau_g)$'s  are   obtained   from   the
integration    of    the    structure   function$^{18}$,
\begin{equation}
n_q=2A\times\frac{\int_{x_{min}}^1u_q(x)dx}{\pi R_A^2\tau_g}
\end{equation}
where  $A$  is the  mass  number  of  the  colliding  nuclei  (208 in
our case for Pb), $x_{min}$ is the minimum value of  $x$, taken$^{1,5}$ to be
0.02, dictated by the applicability of perturbative QCD.
It is well known that the structure functions  depend on
$x$ and $Q^2$ and therefore, it is important to choose the proper range
of $Q^2$. The range of $Q^2$ appropriate for our present context can be
determined by comparing the values of the running coupling constant
$\alpha_s(Q^2)$ with $\alpha_s(T=T_g)$$^{19}$ . Thus for initial
temperatures $T_g$ of 500 and 660 MeV (corresponding to $\tau_g$ = 0.3 and
0.25 fm/c)
at RHIC and LHC energies$^2$, respectively, the appropriate ranges of $Q^2$ are
25 and 55 GeV$^2$. Then, the values of
$n_q(\tau_g)$ estimated from eq. (9) come out to be 1.26 and 1.33
fm$^{-3}$
 at RHIC and LHC
energies. From ref. 16, these values are 0.7 and 2.8 fm$^{-3}$, respectively.

Eqs.(3), (4) and (8) were solved self consistently 
by the Runge~-Kutta method~ to obtain
the~ non-equilibrium density~ $n_q(\tau)$. As is ~evident from
the above discussions, the non-equilibrium density
$n_q$ has an explicit dependence on $\tau$ and an implicit dependence
on $\tau$ through $T(\tau)$. But the equilibrium density $n_{eq}(T)$
has only an implicit dependence on $\tau$ through $T(\tau)$. The
ratio $r_q$ thus assumes an {\it universal} feature, since the implicit
time dependence gets eliminated. The time dependence of the ratio $r_q$ can
then be used as a ready marker for chemical equilibrium; the time at which
the explicit time dependence of ${r_q}$ vanishes, {\it simultaneously with
$r_q \rightarrow$ 1}, corresponds to the time
for chemical equilibration for the flavour $q$.

In contrast to the earlier work of Bir\'o {\it et al}$^{16}$ , we have
not included reactions like ${q\bar q\rightarrow Q\bar Q}$, since
the initial system is dominated by gluons$^{2,7}$;
the quark density is very low compared to the gluons. These
reactions are suppressed by a factor of about ${1/16}$
compared to the case when the quarks are in complete equilibrium$^5$.
Furthermore, we would like to emphasize at this juncture that we are
looking at an optimistic scenario, where the gluons have
completely equilibrated already.

The relative importance of the gluon fusion $(gg\rightarrow q\bar q)$
and gluon decay $(g\rightarrow q \bar q)$ channels is obviously
a crucial issue in the present context. In fig. 1, we show the reaction
rates for these two channels. If one considers the thermal contribution
to the quark masses also, then the effective quark mass becomes
$m_q^{eff}(T)=\sqrt{m_{current}^2+m_{th}^2}$. One can readily see
from fig. 1 that the gluon fusion dominates over the gluon decay modes
for the entire range of interest; it has however to be noticed that
the higher the $m_q^{eff}$ the greater the difference between them,
implying that for light quarks(u,d and even s), the decay channel
does play an important role.

The ratios $r_q=n_q/n_{eq}$ are plotted in fig.2(a) for RHIC
energies for the initial conditions specified by eq. (9). At these
energies, the ratios for $u(d)$ quarks and $s$
quarks are the same at early times but at late times the $u(d)$ quarks
dominate over the $s$ quarks. The reason is that at early times the
thermal mass $(\propto T)$ dominates over the current quark mass for
all flavours. At later times, however, the thermal mass decreases with
decreasing temperature and the difference in the current masses of
$u(d)$ ($\sim 10 MeV$) and $s$ ($\sim 150 MeV$) quarks becomes
important.
There occurs thus some Boltzmann suppression for $s$ quarks
relative to $u(d)$ quarks.

The same calculation has also been carried out for the initial condition
specified by the algorithm HIJING. At RHIC, the initial $r_q$
for all three flavours are lower than those estimated from eq.(9) 
(fig. 2(a)).
As is to be expected, this makes equilibrium even less likely, which
is borne out in our detailed calculation. For LHC, the initial $r_q$'s
are systematically higher than those obtained from eq.(9); 
one could thus naively expect
that equilibration may become easier. We therefore show in fig. 2 (b)
the result for this case also. One sees no sign of saturation with
$\tau$ for either u(d) or s quarks.

In fact, one can also estimate the initial quark density from the prescription
$N_q=\sigma_q\cdot T_{AA}(0)$ where $T_{AA}(0)$ is the overlap function
for central $AA$ collision and $\sigma_q$ the production cross section for the
quark$^{20}$ $q$ , one would obtain significantly
lower values of $n_q(\tau_g)$, or equivalently $r_q(\tau_g)$. ({\it e.g.\/}
$n_q(\tau_g)\sim 0.48$ at RHIC compared to 1.26 obtained from eq.(9)).
This would then imply lesser probability for chemical equilibrium.
We have verified that this is indeed so; the qualitative behaviour
of the results shown above remain unaltered while quantitatively
the conclusions are strengthened at all the energy domains considered here.

We have so far considered futuristic scenarios at RHIC and LHC where
the assumption of a baryon free region is perhaps applicable. For
the presnt, namely the SPS energies, such an approximation is not at all
appropriate; it is expected that there would be a substantial amount
of stopping at these energies$^{21}$. Nevertheless,
study of flavour equilibration at SPS energies becomes a very pertinent and
timely issue. We have therefore investigated two extreme scenarios--
one of total transparency as above and one of complete stopping.

In fig. 3 we show the time evolution of $r_q$ and $T$ as a function of
$\tau$ for the transparent case. The initial conditions are determined
by eq.(9) alone, as, to our knowledge, the corresponding estimates from
HIJING do not exist in the literature.
We find that at early
times $r_s$ is slightly larger than $r_u$ or $r_d$; this is due to
the initial normalisation. The non-equilibrium density $n_{u(d)} \sim n_s$
but the equilibrium density for s quarks is less than that of u or d
quarks because of the difference in their masses. This results
in $r_s$ being larger than $r_{u(d)}$. Expectedly, this effect
washes out with progress of time. In any case, there is no hint
of flavour equilibration at SPS energies either, under the assumption
of complete transparency. In the other extreme case, that of complete
stopping , however, the situation is most interesting. Because of the
presence of a large number of valence u and d quarks, the initial
values of $r_u$ and $r_d$ are very large. For central collisions
of lead nuclei, the initial value of $r_{u(d)}$ turns out to be
$0.98(1.06)$, taking into account both valence and sea contributions.
As $\tau$ increases, $r_q$ tends to 1, $r_u$ from below and $r_d$
from above. Nevertheless,  the stationarity of $r_q$ with $\tau$
does not seem to be achievable within the life time of the QGP.
It is thus fair to conclude that at SPS, u and d quarks, although
not in equilibrium, may not be too far from equilibrium in the
event of complete stopping. The situation for the s quark
is, however, quite different as all the initial s quarks
come from the sea. Thus $r_s(\tau_g)$ is still rather small.
But now, one has to include $u\bar u (d\bar d)\rightarrow s\bar s$,
in addition to gluon fusion and decay channels.
We observe that even for such a favourable
situation the time evolution of $r_s$ is indistinguishable
from that shown in fig. 3.

These findings are obviously in contrast with the earlier expectations
of other authors$^{22,23}$ where s quarks were expected
to come to equilibrium at SPS energies. The difference is ascribable
to the effect of the thermal masses, which must be included
if proper account of the in-medium effects is to be taken. We have
shown here that these effects do play a very important role indeed
and may affect the conclusions nontrivially.
Obviously, implications of these issues in QGP diagnostics
is an urgent task; work along these lines is in progress.
\newpage
\noindent{\large {\bf References}}
\begin{enumerate}

\item J. Alam, S. Raha and B. Sinha, Phys. Rev. Lett.
{\bf 73} (1994) 1895

\item E. Shuryak, Phys. Rev. Lett. {\bf 68} (1992) 3270

\item R. Cutler and D. Sivers, Phys. Rev. {\bf D17} (1978)
196

\item S. Raha, Phys. Scr. {\bf T32} (1990) 180 ; S.
Chakrabarty, S. Raha and B. Sinha, Mod. Phys. Lett. {\bf A7} (1992)
927

\item E. Shuryak and L. Xiong, Phys. Rev. Lett. {\bf 70}
(1993) 2241

\item K. Geiger and J. Kapusta, Phys. Rev. Lett. {\bf 70},
(1993) 1920; Phys. Rev. {\bf D47} (1993) 4905

\item K. Geiger, Phys. Rep. {\bf 258} (1995) 237

\item E. W. Kolb and S. Wolfram, Nucl. Phys. {\bf B172}
(1980) 224

\item E. W. Kolb and M. S. Turner, {\it The Early Universe},
(Addison-Wesley, New York), 1990.

\item J. D. Bjorken, Phys. Rev. {\bf D27} (1993) 140

\item T. Altherr and D. Seibert, Phys. Rev. {\bf C49} (1994)
1684

\item T. Altherr and D. Seibert, Phys. Lett. {\bf B 313}
(1993) 149

\item C. T. Traxler and M. H. Thoma, Phys. Rev. {\bf C53}
(1996)1348.

\item J. Alam {\it et al}, to be published.

\item R. C. Hwa and K. Kajantie, Phys. Rev. {\bf D32} (1985) 1109\\

\item T. S. Bir\'o, E. van Doorn, B. M\"uller, M. H. Thoma and X. N.
Wang, Phys. Rev. {\bf C48} (1993) 1275

\item X. N. Wang, Phys. Rev. {\bf D43} (1991) 104; X. N. Wang
and M. Gyulassy, Phys. Rev. {\bf D44} (1992) 3501

\item M. Gl\"uck, E. Reya and A. Vogt, Z. Phys. {\bf C48} (1990)
471

\item F. Karsch, Z. Phys.  {\bf C38} (1988) 147

\item B. M\"uller and X. N. Wang, Phys. Rev. Lett. {\bf 68}
(1992) 2437

\item F. Plasil in Physics and Astrophysics of Quark-Gluon
Plasma, eds. B. Sinha,
Y. P. Viyogi and S. Raha, (World Scientific, Singapore, 1994).

\item T. Matsui, B. Svetitsky and L. D. Mclerran,
Phys. Rev. {\bf D34}, (1986) 783.

\item J. Rafelski, J. Letessier and A. Tounsi,
in Strangeness in Hadronic Matter, edited by J. Rafelski,
American Institute of Physics, NY, 1995.
\end{enumerate}

\newpage
\section*{Figure Captions}

\noindent Fig.~1 : Comparison of quark production rates
due to gluon fusion and gluon decay, the value of $\alpha_s$
is calculated from the parametrisation of ref 19.
\vskip 0.4 cm

\noindent Fig.~2 : Ratio of non-equilibrium density to equilibrium density
and temperature as functions of time (a) at RHIC
energy for initial condition given by eq.(9)
and (b) at LHC energy for intial condition given by eq.(9)
and ref. 16.
\vskip .4 cm
\noindent Fig.~3 : Same as fig.~2 (a), at SPS energy (see text).
\newpage
Running title:

{\Large {Thermal Masses and Equilibrium Rates .....}}

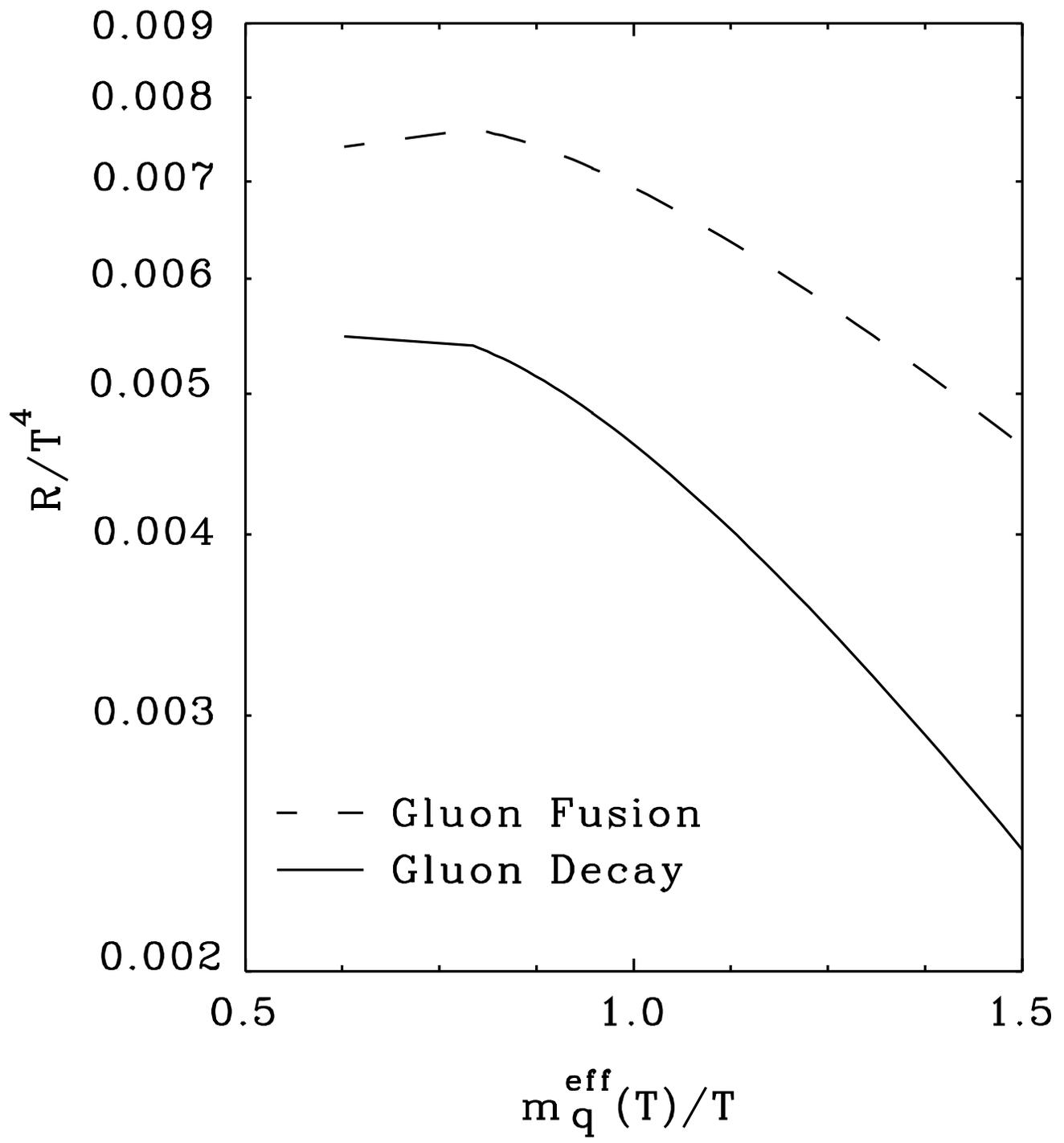
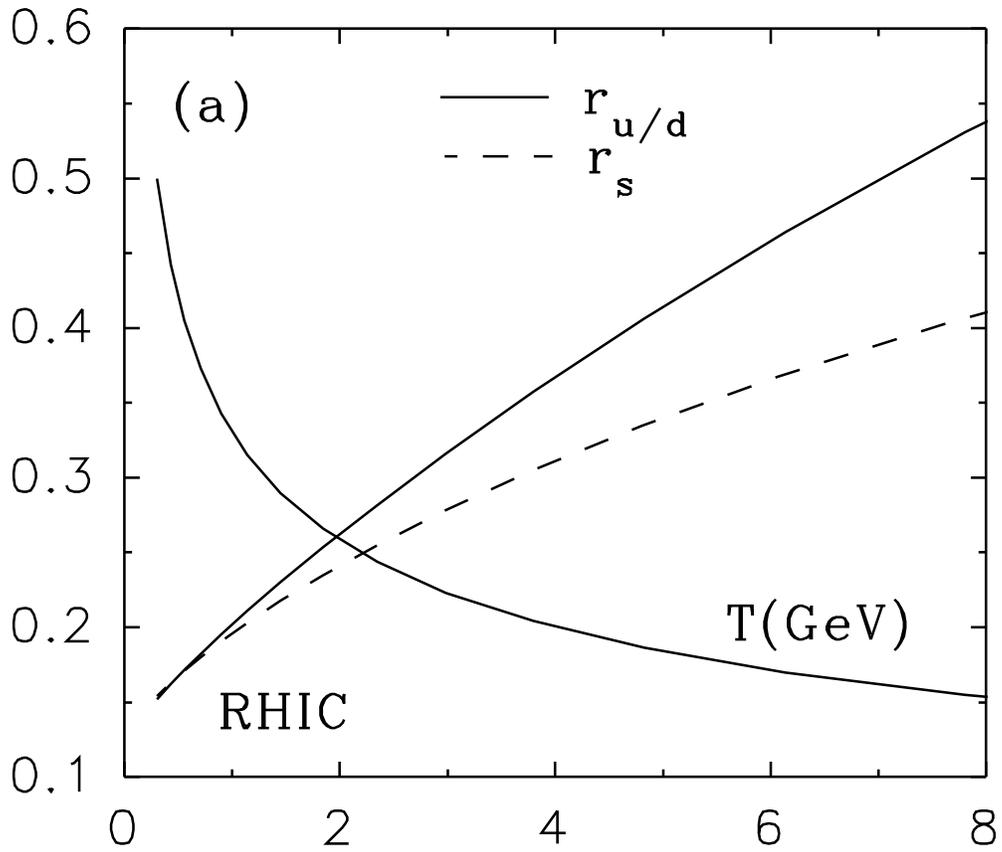
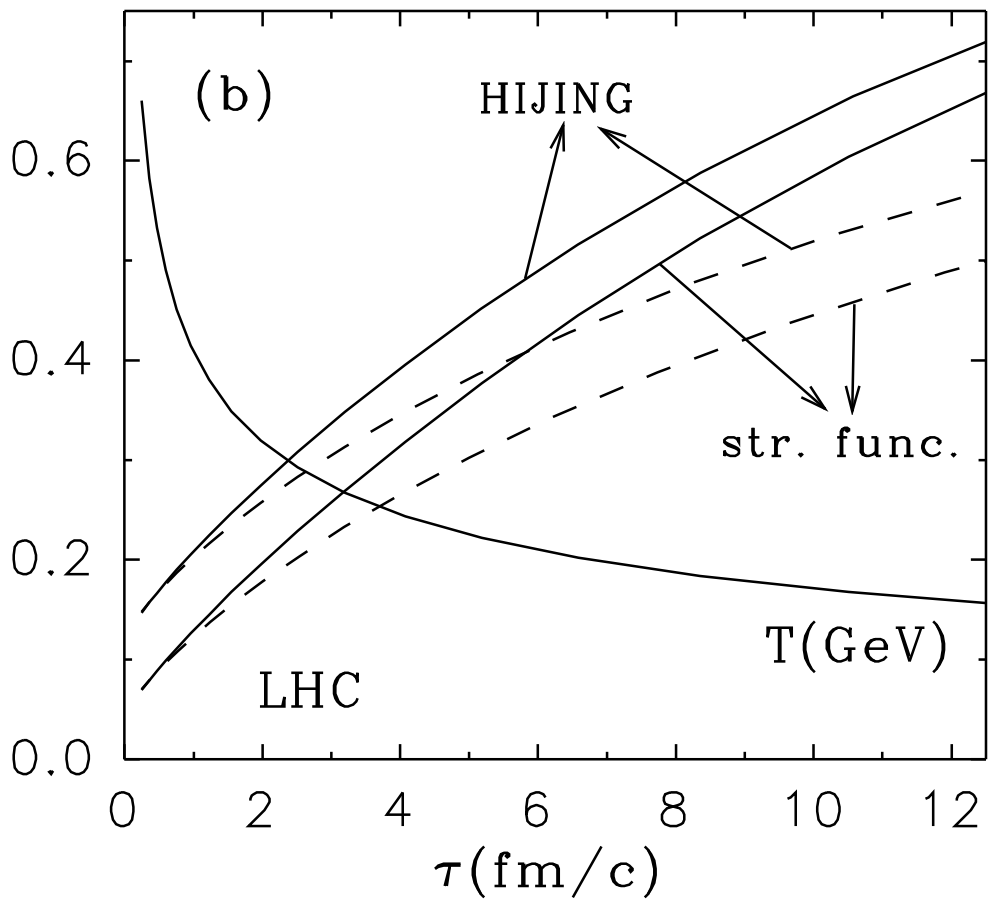
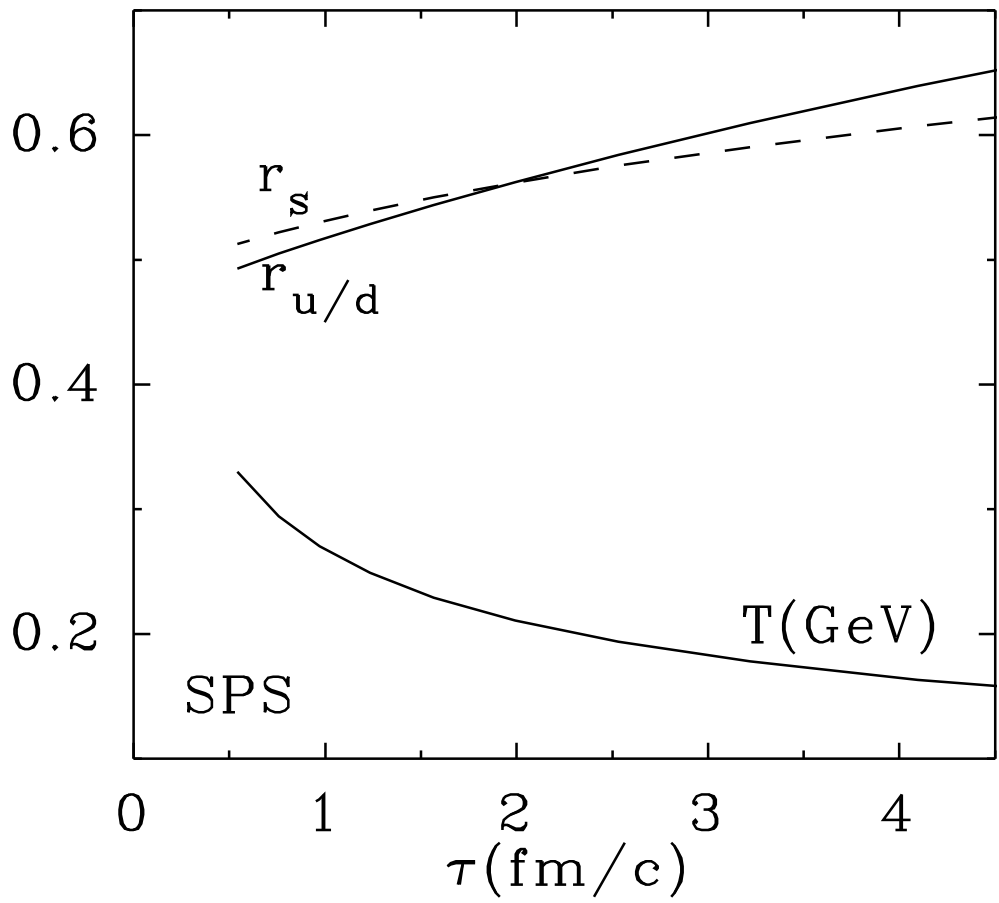
\end{document}